\documentclass[prb,aps,twocolumn,showpacs]{revtex4}

\usepackage{graphicx,latexsym}
\usepackage{dcolumn}
\usepackage{amsmath,amssymb,epsf,bm}

\begin{document}

\title{Electrical conductivity beyond linear response in layered
superconductors under magnetic field}
\author{Bui Duc Tinh$^{1}$, Dingping Li$^{2}$, and Baruch Rosenstein$^{1}$}
\affiliation{$^{1}$Department of Electrophysics, National Chiao Tung University, Hsinchu
30050, Taiwan, Republic of China\\
$^{2}$Department of Physics, Peking University, Beijing 100871, China}

\begin{abstract}
The time-dependent Ginzburg-Landau approach is used to investigate nonlinear
response of a strongly type-II superconductor. The dissipation takes a form
of the flux flow which is quantitatively studied beyond linear response.
Thermal fluctuations, represented by the Langevin white noise, are assumed
to be strong enough to melt the Abrikosov vortex lattice created by the
magnetic field into a moving vortex liquid and marginalize the effects of
the vortex pinning by inhomogeneities. The layered structure of the
superconductor is accounted for by means of the Lawrence-Doniach model. The
nonlinear interaction term in dynamics is treated within Gaussian
approximation and we go beyond the often used lowest Landau level
approximation to treat arbitrary magnetic fields. The I-V curve is
calculated for arbitrary temperature and the results are compared to
experimental data on high-$T_{c}$ superconductor YBa$_{2}$Cu$_{3}$O$%
_{7-\delta }$.
\end{abstract}

\maketitle

\section{Introduction}

Electric response of a high temperature superconductor (HTSC) under magnetic
field has been a subject of extensive experimental and theoretical
investigation for years. Magnetic field in these layered strongly type-II
superconductors create magnetic vortices, which, if not pinned by
inhomogeneities, move and let the electric field to penetrate the mixed
state. The dynamic properties of fluxons appearing in the bulk of a sample
are strongly affected by the combined effect of thermal fluctuations,
anisotropy (dimensionality) and the flux pinning \cite{Blatter}. Thermal
fluctuations in these materials are far from negligible and in particular
are responsible for existence of the first-order vortex lattice melting
transition separating two thermodynamically distinct phases, the vortex
solid and the vortex liquid. Magnetic field and reduced dimensionality due
to pronounced layered structure (especially in materials like Bi$_{2}$Sr$%
_{2} $CaCuO$_{8+\delta }$) further enhance the effect of thermal
fluctuations on the mesoscopic scale. On the other hand the role of pinning
in high-$T_{c}$ materials is reduced significantly compared to the low
temperature one, leading to smaller critical currents. At elevated
temperatures the thermal depinning \cite{Blatter} further diminishes effects
of disorder.

Linear response to electric field in the mixed state of these
superconductors has been thoroughly explored experimentally and
theoretically over the last three decades. These experiments were performed
at very small voltages in order to avoid effects of nonlinearity. Deviation
from linearity however are interesting in their own right. These effects
have also been studied in low-$T_{c}$ superconductors experimentally\cite%
{Kajimura71,Grenoble} and theoretically \cite{Schmid,Li04}\ and
recently experiments were extended to HTSC compounds
\cite{Livanov,Puica09}.

Since thermal fluctuations in the low-$T_{c}$ materials are
negligible compared to the inter-vortex interactions, the moving
vortex matter is expected to preserve a regular lattice structure
(for weak enough disorder). On the other hand, as mentioned above,
the vortex lattice melts in HTSC over large portions of their phase
diagram, so the moving vortex matter in the region of vortex liquid
can be better described as an irregular flowing vortex liquid. In
particular the nonlinear effects will also be strongly influenced by
the thermal fluctuations.

A simpler case of a zero or very small magnetic field in the case of strong
thermal fluctuations was in fact comprehensively studied theoretically \cite%
{Varlamov} albeit in linear response only. In any superconductor there
exists a critical region around the critical temperature $\left\vert
T-T_{c}\right\vert \ll Gi\cdot T_{c}$, in which the fluctuations are strong
(the Ginzburg number characterizing the strength of thermal fluctuations is
just $Gi\sim 10^{-10}-10^{-7}$ for low $T_{c}$, while $Gi\sim
10^{-5}-10^{-1} $ for HTSC materials). Outside the critical region and for
small electric fields, the fluctuation conductivity was calculated by
Aslamazov and Larkin \cite{Larkin68} by considering (noninteracting)
Gaussian fluctuations within Bardeen-Cooper-Schrieffer (BCS) and within a
more phenomenological Ginzburg-Landau (GL) approach. In the framework of the
GL approach (restricted to the lowest Landau level approximation), Ullah and
Dorsey \cite{Ullah90} computed the Ettingshausen coefficient by using the
Hartree approximation. This approach was extended to other transport
phenomena like the Hall conductivity \cite{Ullah90} and the Nernst effect
\cite{Tinh}.

The fluctuation conductivity within linear response can be applied to
describe sufficiently weak electric fields, which do not perturb the
fluctuations' spectrum \cite{Hurault69}. Physically at electric field, which
is able to accelerate the paired electrons on a distance of the order of the
coherence length $\xi \,$ so that they change their energy by a value
corresponding to the Cooper pair binding energy, the linear response is
already inapplicable \cite{Varlamov}. The resulting additional field
dependent depairing leads to deviation of the current-voltage
characteristics from the Ohm's law. The non-Ohmic fluctuation conductivity
was calculated for a layered superconductor in an arbitrary electric field
considering the fluctuations as noninteracting Gaussian ones \cite%
{Varlamov92,Mishonov02}. The fluctuations' suppression effect of high
electric fields in HTSC was investigated experimentally for the in-plane
paraconductivity in zero magnetic field \cite{Soret93,Fruchter04,Lang05},
and a good agreement with the theoretical models \cite{Varlamov92,Mishonov02}
was found.

In this paper the nonlinear electric response of the moving vortex liquid in
a layered superconductor under magnetic field perpendicular to the layers is
studied using the time dependent GL (TDGL) approach. The layered structure
is modeled via the Lawrence-Doniach discretization in the magnetic field
direction. In the moving vortex liquid the long range crystalline order is
lost due to thermal fluctuations and the vortex matter becomes homogeneous
on a scale above the average inter-vortex distances. Although sometimes
motion tends to suppress the fluctuations, they are still a dominant factor
in flux dynamics. The TDGL approach is an ideal tool to study a combined
effect of the dissipative (overdamped) flux motion and thermal fluctuations
conveniently modeled by the Langevin white noise. The interaction term in
dynamics is treated in Gaussian approximation which is similar in structure
to the Hartree-Fock one.

Theoretically the nonlinear effects in HTSC have been addressed
\cite{Puica03}. However the results of Ref. \onlinecite{Puica03} are
different from our in this paper and we will sketch the difference
below.

Firstly the model of Ref. \onlinecite{Puica03}, is physically
different from ours. The authors in Ref. \onlinecite{Puica03}
believe that the two quantities, layer distance and thickness in the
Lawrence-Doniach for HTSC are equal (apparently not the case in
HTSC), while we consider them as two independent parameters. Another
difference is we use so called self-consistent Gaussian
approximation to treat the model while Ref. \onlinecite{Puica03}
used the Hartree-Fock approximation.

A main contribution of our paper is an explicit form of the Green
function incorporating all Landau levels. This allows to obtain
explicit formulas without need to cutoff higher Landau levels. In
Ref. \onlinecite{Puica03}, a nontrivial matrix inversion (of
infinite matrices) or cutting off the number of Landau levels is
required. Note that the exact analytical expression of Green
function of the linearized TDGL equation in DC field can be even
generalized also to AC field. The method is very general, and it
allow us to study transport phenomena beyond linear response of
type-II superconductor like the Nernst effect, Hall effect. The
renormalization of the models is also different from Ref.
\onlinecite{Puica03}. One of the main result of our work is that the
conductivity formula is independent of UV cutoff (unlike in Ref.
\onlinecite{Puica03}) as it should be as the standard $|\Psi |^{4}$
theory is renormalizable. Furthermore Gaussian approximation used in
this paper is consistent to leading order with perturbation theory,
see Ref. \onlinecite{Kovner89} in which it is shown that this
procedure preserved a correct the ultraviolet (UV) renormalization
(is RG invariant). Without electric field the issue was
comprehensively discussed in a textbook Kleinert \cite{Kovner89}.
One can use Hartree-Fock procedure only when UV issues are
unimportant. We can also show, if there is no electric field, the
result obtained using TDGL model and Gaussian approximation will
lead the same thermodynamic equation using Gaussian approximation.

The paper is organized as follows. The model is defined in Sec. II. The
vortex liquid within the Gaussian approximation is described in Sec. III.
The I-V curve and the comparison with experiment are described in Sec. IV,
while Sec. V contains conclusions.

\section{Thermal fluctuations in the time dependent GL Lawrence-Doniach model%
}

To describe fluctuation of order parameter in layered superconductors, one
can start with the Lawrence-Doniach expression of the GL free energy of the
2D layers with a Josephson coupling between them:
\begin{eqnarray}
F_{GL} &=&s^{\prime }\sum_{n}\int d^{2}r\Big\{\frac{{\hbar }^{2}}{2m^{\ast }}%
|\mathbf{D}\Psi _{n}|^{2}+\frac{{\hbar }^{2}}{2m_{c}d^{\prime 2}}|\Psi
_{n}-\Psi _{n+1}|^{2}  \notag \\
&+&a|\Psi _{n}|^{2}+\frac{b^{\prime }}{2}|\Psi _{n}|^{4}\Big\},
\label{original.Fre}
\end{eqnarray}%
where $s^{\prime }$ is the order parameter effective
\textquotedblleft thickness" and $d^{\prime }$ distance between
layers labeled by $n$. The Lawrence-Doniach model approximates
paired electrons DOS by homogeneous infinitely thin planes separated
by distance $d^{\prime }$. While discussing thermal fluctuations, we
have to introduce a finite thickness, otherwise the fluctuations
will not allow the condensate to exist (Mermin-Wagner theorem). The
thickness is of course smaller than the distance between the layers
(otherwise we would not have layers). The order parameter is assumed
to be non-zero within $s^{\prime }$. Effective Cooper pair mass in
the $ab$ plane is $m^{\ast }$(disregarding for simplicity the
anisotropy between the crystallographic $a$ and $b$ axes), while
along the $c$ axis it is much
larger-$m_{c}$. For simplicity we assume $a=\alpha T_{c}^{mf}(t-1)$, $%
t\equiv T/T_{c}^{mf}$, although this temperature dependence can be
easily modified to better describe the experimental coherence
length. The \textquotedblleft mean field" critical temperature
$T_{c}^{mf}$ depends on UV cutoff, $\tau _{c}$, of the
\textquotedblleft mesoscopic" or \textquotedblleft phenomenological"
GL description, specified later. This temperature is higher than
measured critical temperature $T_{c}$ due to strong thermal
fluctuations on the mesoscopic scale.

The covariant derivatives are defined by $\mathbf{D}\equiv \mathbf{\nabla }%
+i(2\pi /\Phi _{0})\mathbf{A},$ where the vector potential describes
constant and homogeneous magnetic field $\mathbf{A}=\mathbf{(-}By,0\mathbf{)}
$ and $\Phi _{0}=hc/e^{\ast }$ is the flux quantum with $e^{\ast
}=2\left\vert e\right\vert $. The two scales, the coherence length $\xi ^{2}=%
{\hbar }^{2}/(2m^{\ast }\alpha T_{c}),$ \ and the penetration depth,\ \ $%
\lambda ^{2}=c^{2}m^{\ast }b^{\prime }/(4\pi e^{\ast 2}\alpha T_{c})$ define
the GL ratio $\kappa \equiv \lambda /\xi $, which is very large for HTSC. In
this case of strongly type-II superconductors the magnetization is by a
factor $\kappa ^{2}$ smaller than the external field for magnetic field
larger than the first critical field $H_{c1}\left( T\right) $, so that we
take $B\approx H$. The electric current, $\mathbf{J}=\mathbf{J}_{n}+\mathbf{J%
}_{s}$, includes both the Ohmic normal part
\begin{equation}
\mathbf{J}_{n}=\sigma _{n}\mathbf{E},  \label{Ohm}
\end{equation}%
and the supercurrent
\begin{equation}
\mathbf{J}_{s}=\frac{ie^{\ast }{\hbar }}{2m^{\ast }}\left( \Psi _{n}^{\ast }%
\mathbf{D}\Psi _{n}-\Psi _{n}\mathbf{D}\Psi _{n}^{\ast }\right) \text{.}
\label{Js_I}
\end{equation}%
Since we are interested in a transport phenomenon, it is necessary to
introduce a dynamics of the order parameter. The simplest one is a
gauge-invariant version of the \textquotedblleft type A" relaxational
dynamics \cite{Ketterson}. In the presence of thermal fluctuations, which on
the mesoscopic scale are represented by a complex white noise \cite%
{Rosenstein07}, it reads:
\begin{equation}
\frac{{\hbar }^{2}\gamma ^{\prime }}{2m^{\ast }}D_{\tau }\Psi _{n}=-\frac{1}{%
s^{\prime }}\frac{\delta F_{GL}}{\delta \Psi _{n}^{\ast }}+\zeta _{n}\text{,}
\label{TDGL_I}
\end{equation}%
where $D_{\tau }\equiv \partial /\partial \tau -i(e^{\ast }/\hbar )\Phi $ is
the covariant time derivative, with $\Phi =-Ey$ being the scalar electric
potential describing the driving force in a purely dissipative dynamics. The
electric field is therefore directed along the $y$ axis and consequently the
vortices are moving in the $x$ direction. For magnetic fields that are not
too low, we assume that the electric field is also homogeneous \cite%
{Rosenstein07}. The inverse diffusion constant $\gamma ^{\prime }/2$,
controlling the time scale of dynamical processes via dissipation, is real,
although a small imaginary (Hall) part is also generally present \cite{Troy}%
. The variance of the thermal noise, determining the temperature $T$ is
taken to be the Gaussian white noise:
\begin{equation}
\left\langle \zeta _{n}^{\ast }(\mathbf{r},\tau )\zeta _{m}(\mathbf{r}%
^{\prime },\tau ^{\prime })\right\rangle =\frac{{\hbar }^{2}\gamma ^{\prime }%
}{m^{\ast }s^{\prime }}T\delta (\mathbf{r}-\mathbf{r}^{\prime })\delta (\tau
-\tau ^{\prime })\delta _{nm}\text{.}  \label{noise}
\end{equation}

Throughout most of the paper we use the coherence length $\xi $ as a unit of
length and $H_{c2}=\Phi _{0}/2\pi \xi ^{2}$ as a unit of the magnetic field.
The dimensionless Boltzmann factor in these units is:
\begin{eqnarray}
\frac{F_{GL}}{T} &=&\frac{s}{\omega t}\sum_{n}\int d^{2}r\Big\{\frac{1}{2}%
|D\psi _{n}|^{2}+\frac{1}{2d^{2}}|\psi _{n}-\psi _{n+1}|^{2}  \notag \\
&-&\frac{1-t}{2}|\psi _{n}|^{2}+\frac{1}{2}|\psi _{n}|^{4}\Big\},
\label{Boltz}
\end{eqnarray}%
where the covariant derivatives in dimensionless units in Landau gauge are $%
D_{x}=\frac{\partial }{\partial x}-iby,$ $D_{y}=\frac{\partial }{\partial y}$
with $b=B/H_{c2}$, and the order parameter field was rescaled: $\Psi
^{2}=(2\alpha T_{c}^{mf}/b^{\prime })\psi ^{2}$. The dimensionless
fluctuations' strength coefficient is
\begin{equation}
\omega =\sqrt{2Gi}\pi ,  \label{omega}
\end{equation}%
where the Ginzburg number is defined by

\begin{equation}
Gi=\frac{1}{2}\left( 8e^{2}\kappa ^{2}\xi T_{c}^{mf}\gamma /c^{2}{\hbar }%
^{2}\right) ^{2}\text{.}  \label{Gno}
\end{equation}%
Note that here we use the standard definition of the Ginzburg number
different from that in Ref. \onlinecite{Li01}. The relation between
parameters of the two models, the Lawrence-Doniach and the 3D anisotropic GL
model, is $d^{\prime }=d\xi _{c}=d\xi /\gamma $, $s^{\prime }=s\xi _{c}=s\xi
/\gamma $, where $\gamma ^{2}\equiv m_{c}/m^{\ast }$ is an anisotropy
parameter. In analogy to the coherence length and the penetration depth, one
can define a characteristic time scale. In the superconducting phase a
typical \textquotedblleft relaxation" time is $\tau _{GL}=\gamma ^{\prime
}\xi ^{2}/2$. It is convenient to use the following unit of the electric
field and the dimensionless field: $E_{GL}=H_{c2}\xi /c\tau _{GL},\mathcal{E}%
=E/E_{GL}$. The TDGL Eq. (\ref{TDGL_I}) written in dimensionless units reads
\begin{equation}
\widehat{H}\psi _{n}+\frac{1}{2d^{2}}(2\psi _{n}-\psi _{n+1}-\psi _{n-1})-%
\frac{1-t}{2}\psi _{n}+|\psi _{n}|^{2}\psi _{n}=\overline{\zeta _{n}},
\label{TDGL2l}
\end{equation}%
\begin{equation*}
\widehat{H}=D_{\tau }-\frac{1}{2}D^{2},
\end{equation*}%
while the Gaussian white-noise correlation takes a form
\begin{equation}
\left\langle \overline{\zeta _{n}}^{\ast }(\mathbf{r},\tau )\overline{\zeta
_{m}}(\mathbf{r}^{\prime },\tau ^{\prime })\right\rangle =\frac{2\omega t}{s}%
\delta (\mathbf{r}-\mathbf{r}^{\prime })\delta (\tau -\tau ^{\prime })\delta
_{nm}\text{.}  \label{correlation_noise}
\end{equation}%
The covariant time derivative in dimensionless units is $D_{\tau }=\frac{%
\partial }{\partial \tau }+ivby$ with $v=\mathcal{E}/b$ being the vortex
velocity and the thermal noise was rescaled as $\zeta _{n}$ $=\overline{%
\zeta _{n}}(2\alpha T_{c}^{mf})^{3/2}/b^{\prime 1/2}$. The dimensionless
current density is $\mathbf{J}_{s}=J_{GL}j_{s}$ where
\begin{equation}
\text{\ \ \ \ \ }j_{s}=\frac{i}{2}\left( \psi _{n}^{\ast }D\psi _{n}-\psi
_{n}D\psi _{n}^{\ast }\right) .  \label{current}
\end{equation}%
with $J_{GL}=cH_{c2}/(2\pi \xi \kappa ^{2})$ being the unit of the current
density. Consistently the conductivity will be given in units of $\sigma
_{GL}=J_{GL}/E_{GL}=c^{2}\gamma ^{\prime }/(4\pi \kappa ^{2})$. This unit is
close to the normal state conductivity $\sigma _{n}$ in dirty limit
superconductors \cite{Kopnin}. In general there is a factor $k$ of order one
relating the two: $\sigma _{n}=k\sigma _{GL}$.

\section{Vortex liquid within the gaussian approximation}

\subsection{Gap equation}

Thermal fluctuations in vortex liquid frustrate the phase of the order
parameter, so that $\langle \psi _{n}(\mathbf{r},\tau )\rangle =0$.
Therefore the contributions to the expectation values of physical quantities
like the electric current come exclusively from the correlations, the most
important being the quadratic one $\left\langle \psi _{n}(\mathbf{r},\tau
)\psi _{n}^{\ast }(\mathbf{r}^{\prime },\tau ^{\prime })\right\rangle $. In
particular, $\langle |\psi _{n}(\mathbf{r},\tau )|^{2}\rangle $ is the
superfluid density. A simple approximation which captures the most
interesting fluctuations effects in the Gaussian approximation, in which the
cubic term in the TDGL Eq. (\ref{TDGL2l}), $|\psi _{n}|^{2}\psi _{n}$, is
replaced by a linear one $2\left\langle |\psi _{n}|^{2}\right\rangle \psi
_{n}$
\begin{equation}
\left( \widehat{H}-\frac{b}{2}\right) \psi _{n}+\frac{1}{2d^{2}}(2\psi
_{n}-\psi _{n+1}-\psi _{n-1})+\varepsilon \psi _{n}=\overline{\zeta _{n}}%
\text{,}  \label{TDGL3}
\end{equation}%
leading the \textquotedblleft renormalized" value of the coefficient of the
linear term:
\begin{equation}
\varepsilon =-a_{h}+2\left\langle |\psi _{n}|^{2}\right\rangle \text{,}
\label{gap.eq1}
\end{equation}%
where the constant is defined as $a_{h}=(1-t-b)/2$. The average $%
\left\langle |\psi _{n}|^{2}\right\rangle $ is expressed via the parameter $%
\varepsilon $ below and will be determined self-consistently together with $%
\varepsilon $. It differs slightly from a well known Hartree-Fock procedure
in which the coefficient of the linearized term is generally different (see
\cite{Kovner89} for details).

Due to the discrete translation invariance in the field direction $z$, it is
convenient to work with the Fourier transform with respect to the layer
index:
\begin{eqnarray}
\psi _{n}\left( \mathbf{r,}\tau \right) &=&\int_{0}^{2\pi /d}\frac{dk_{z}}{%
2\pi }e^{-ink_{z}d}\psi _{k_{z}}(\mathbf{r},\tau )\text{,}  \notag \\
\psi _{k_{z}}\left( \mathbf{r,}\tau \right) &=&d\sum_{n}e^{ink_{z}d}\psi
_{n}(\mathbf{r},\tau )\text{,}  \label{Fourier}
\end{eqnarray}%
and similar transformation for $\overline{\zeta }$. In terms of Fourier
components the TDGL Eq. (\ref{TDGL3}) becomes
\begin{equation}
\left\{ \widehat{H}-\frac{b}{2}+\frac{1}{d^{2}}[1-\cos (k_{z}d)]+\varepsilon
\right\} \psi _{k_{z}}(\mathbf{r},\tau )=\overline{\zeta _{k_{z}}}(\mathbf{r}%
,\tau )\text{.}  \label{TDGL_kz1}
\end{equation}%
The noise correlation is
\begin{equation}
\left\langle \overline{\zeta _{k_{z}}}^{\ast }(\mathbf{r},\tau )\overline{%
\zeta _{k_{z}^{\prime }}}(\mathbf{r}^{\prime },\tau ^{\prime })\right\rangle
=4\pi \omega t\frac{d}{s}\delta (\mathbf{r}-\mathbf{r}^{\prime })\delta
(\tau -\tau ^{\prime })\delta (k_{z}-k_{z}^{\prime })\text{.}
\end{equation}%
The relaxational linearized TDGL equation with a Langevin noise, Eq. (\ref%
{TDGL_kz1}), is solved using the retarded ($G=0$ for $\tau <\tau ^{\prime }$%
) Green function (GF) $G_{k_{z}}(\mathbf{r},\tau ;\mathbf{r}^{\prime },\tau
^{\prime })$:
\begin{eqnarray}
\psi _{n}(\mathbf{r},\tau ) &=&\int_{0}^{2\pi /d}\frac{dk_{z}}{2\pi }%
e^{-ink_{z}d}\int d\mathbf{r}^{\prime }  \notag \\
&&\times \int d\tau ^{\prime }G_{k_{z}}(\mathbf{r},\tau ;\mathbf{r}^{\prime
},\tau ^{\prime })\overline{\zeta _{k_{z}}}(\mathbf{r}^{\prime },\tau
^{\prime }).
\end{eqnarray}%
The GF satisfies
\begin{eqnarray}
&&\left\{ \widehat{H}-\frac{b}{2}+\frac{1}{d^{2}}[1-\cos
(k_{z}d)]+\varepsilon \right\} G_{k_{z}}(\mathbf{r},\mathbf{r}^{\prime
},\tau -\tau ^{\prime })  \notag \\
&=&\delta (\mathbf{r}-\mathbf{r}^{\prime })\delta (\tau -\tau ^{\prime }),
\label{GFdef}
\end{eqnarray}%
and is computed in the Appendix A.

The thermal average of the superfluid density (density of Cooper pairs) is
\begin{eqnarray}
\left\langle \left\vert \psi _{n}(\mathbf{r},\tau )\right\vert
^{2}\right\rangle &=&2\omega t\frac{d}{s}\int_{0}^{2\pi /d}\frac{dk_{z}}{%
2\pi }\int d\mathbf{r}^{\prime }  \notag \\
&&\times \int d\tau ^{\prime }\left\vert G_{k_{z}}(\mathbf{r-r}^{\prime
},\tau -\tau ^{\prime })\right\vert ^{2}  \notag \\
&=&\frac{\omega tb}{2\pi s}\int_{\tau =\tau _{c}}^{\infty }\frac{%
f(\varepsilon ,\tau )}{\sinh (b\tau )},  \label{expect.v}
\end{eqnarray}%
where
\begin{eqnarray}
f(\varepsilon ,\tau ) &=&\exp \left[ \frac{2v^{2}}{b}\tanh \left( \frac{%
b\tau }{2}\right) \right] e^{-\left( 2\varepsilon -b+v^{2}\right) \tau }
\notag \\
&&\times e^{-2\tau /d^{2}}I_{0}\left( 2\tau /d^{2}\right) \text{.}  \label{f}
\end{eqnarray}%
Here $I_{0}(x)=(1/2\pi )\int_{0}^{2\pi }e^{x\cos \theta }d\theta $ is the
modified Bessel function. The first pair of multipliers in Eq. (\ref{f}) is
independent of the inter-plane distance $d$ and exponentially decreases for $%
\tau >\left( 2\varepsilon -b+v^{2}\right) ^{-1}$, while the last pair of
multipliers depends on the layered structure. The expression (\ref{expect.v}%
) is divergent at small $\tau $, so an UV cutoff $\tau _{c}$ is necessary
for regularization. Substituting the expectation value into the
\textquotedblleft gap equation", Eq. (\ref{gap.eq1}), the later takes a form
\begin{equation}
\varepsilon =-a_{h}+\frac{\omega tb}{\pi s}\int_{\tau =\tau _{c}}^{\infty }%
\frac{f(\varepsilon ,\tau )}{\sinh (b\tau )}.  \label{gap.eq2}
\end{equation}

\subsection{Renormalization}

In order to absorb the divergence into a \textquotedblleft renormalized"
value $a_{h}^{r}$ of the coefficient $a_{h}$, it is convenient to make an
integration by parts in the last term for small $\tau _{c}$:
\begin{eqnarray}
&&b\int_{\tau =\tau _{c}}^{\infty }\frac{f(\tau )}{\sinh (b\tau )}  \notag \\
&\simeq &-\int_{0}^{\infty }\ln [\sinh (b\tau )]\frac{d}{d\tau }\left[ \frac{%
f(\varepsilon ,\tau )}{\cosh (b\tau )}\right] -\ln (b\tau _{c}).
\end{eqnarray}%
Physically the renormalization corresponds to reduction of the critical
temperature by the thermal fluctuations from $T_{c}^{mf}$ to $T_{c}$. The
thermal fluctuations occur on the mesoscopic scale. The critical temperature
$T_{c}$ is defined at $\varepsilon =0$, and $\upsilon =0$, and at low
magnetic field less than $H_{c1}=\frac{H_{c2}}{2\kappa ^{2}}\ln \left(
\kappa \right) $ (for a typical high $T_{c}$ superconductor, $\kappa \simeq
50$, $\ H_{c1}=7.8\times 10^{-4}H_{c2}$ ), the superconductor is at Meissner
phase, $b=0$, \ leading to
\begin{equation}
T_{c}=T_{c}^{mf}\left\{ 1+\frac{2\omega }{\pi s}\left[ \ln \left( \tau
_{c}/d^{2}\right) +\gamma _{E}\right] \right\} ,  \label{Tmf}
\end{equation}%
where $\gamma _{E}=0.577$ is Euler constant, and Eq. (\ref{gap.eq2}) can be
rewritten as
\begin{eqnarray}
\varepsilon  &=&-a_{h}^{r}-\frac{\omega t}{\pi s}\int_{0}^{\infty }\ln
[\sinh (b\tau )]\frac{d}{d\tau }\left[ \frac{f(\varepsilon ,\tau )}{\cosh
(b\tau )}\right]   \notag \\
&&+\frac{\omega t}{\pi s}\left\{ \gamma _{E}-\ln (bd^{2})\right\} \text{,}
\label{gapequation}
\end{eqnarray}%
where $a_{h}^{r}=\frac{1-b-T/T_{c}}{2}$, $t=T/T_{c}$, and $\omega
=\sqrt{2Gi}\pi$ where $Gi=\frac{1}{2}\left( 8e^{2}\kappa ^{2}\xi
T_{c}\gamma /c^{2}{\hbar }^{2}\right) ^{2}$($T_{c}^{mf}$ is now
replaced by $T_{c}$). The formula is cutoff independent. In terms of
energy UV cutoff $\Lambda $, introduced for example in \cite{Tinh},
the cutoff \textquotedblleft time\textquotedblright\ $\tau _{c}$ can
be expressed as

\begin{equation}
\tau _{c}=1/(2e^{\gamma _{E}}\Lambda ).  \label{tcutoff}
\end{equation}%
This is obtained by comparing a thermodynamic result for a physical
quantity like superfluid density with the dynamic result (see
Appendix B). The temporary UV cutoff used is completely equivalent
to the standard energy or momentum cutoff Lambda used in
thermodynamics (in which the time dependence does not appear).
Physically one might think about momentum cutoff as more basic and
this would be universal and independent of particular time dependent
realization of thermal fluctuations (TDGL with white noise in our
case). Roughly (in physical units) $\Lambda \simeq \varepsilon
_{F}=\hbar ^{2}k_{F}^{2}/(2m^{\ast })$. In the next section we will
discuss the estimate of $T_{c}^{mf}$ using this value due to the
following reason. For high-$T_{c}$ materials ordinary BCS is invalid
and coherence length is of order of lattice spacing (the cutoff
becomes microscopic) and therefore the energy cutoff is of order
$\varepsilon _{F}$. Except the formula to calculate $T_{c}^{mf}$,
all other formulas in this paper is independent of energy cutoff.

\section{The I-V curve}

\subsection{Current density}

The supercurrent density, defined by Eq. (\ref{current}), can be expressed
via the Green's functions as:
\begin{eqnarray}
j_{y}^{s} &=&i\omega t\frac{d}{s}\int_{0}^{2\pi /d}\frac{dk_{z}}{2\pi }%
\int_{r^{\prime },\tau ^{\prime }}G_{k_{z}}^{\ast }\left( \mathbf{r}-\mathbf{%
r}^{\prime },\tau -\tau ^{\prime }\right)  \notag \\
&&\times \frac{\partial }{\partial y}G_{k_{z}}\left( \mathbf{r}-\mathbf{r}%
^{\prime },\tau -\tau ^{\prime }\right) +c.c.
\end{eqnarray}%
Performing the integrals, one obtains:
\begin{equation}
j_{y}^{s}=\frac{\omega t}{4\pi s}\nu \int_{\tau =0}^{\infty }\frac{%
f(\varepsilon ,\tau )}{\cosh (\frac{\tau }{2})^{2}}\text{,}
\end{equation}%
where the function $f$ was defined in Eq. (\ref{f}). Consequently the
contribution to the conductivity is $\overline{\sigma} ^{s}=j_{y}^{s}/%
\mathcal{E}$. The conductivity expression (Eq. 27) is not divergent when
expressed as a function of renormalized $T_{c}$ (the real transition
temperature), so it is independent of the cutoff. This is considered in
detail in section III. B and is indeed different from the Ref. %
\onlinecite{Puica03}. In physical units the current density reads
\begin{equation}
J_{y}=\sigma _{n}E\left[ 1+\frac{\omega t}{4\pi s}\frac{1}{k}\int_{\tau
=0}^{\infty }\frac{f(\varepsilon ,\tau )}{\cosh (\frac{b\tau }{2})^{2}}%
\right] .  \label{finalcurrent}
\end{equation}%
This is the main result of the present paper. We also considered the
conductivity expression in 2D in linear response which do match the linear
response conductivity expression derived in our previous work \cite{Tinh}.
\begin{equation}
\overline{\sigma} ^{s}_{2D} =\frac{\omega t}{4\pi s b}\left\{ 2-\left( 1-%
\frac{2\varepsilon }{b}\right) \left[ \psi\left( \frac{\varepsilon }{b}%
\right) -\psi\left( \frac{1}{2}+\frac{\varepsilon }{b}\right) \right]
\right\} ,  \label{conduc2Dlinear}
\end{equation}
where $\psi$ is the polygamma function.

\subsection{Comparison with experiment}

In this section we use physical units, while the dimensionless quantities
are denoted with bars. The experiment results of I. Puica \textit{et al.}%
\cite{Puica09}, obtained from the resistivity and Hall effect
measurements on an optimally doped YBa$_{2}$Cu$_{3}$O$_{7-\delta }$
(YBCO) films of thickness $50$ nm and $T_{c}=86.8$ K. The distance
between the bilayers used the calculation is $d^{\prime }=11.68$
\AA\ in Ref. \onlinecite{Yan}. The number of bilayers is $50$, large
enough to be described by the Lawrence-Doniach model without taking
care of boundary conditions. In order to compare the fluctuation
conductivity with experimental data in HTSC, one cannot use the
expression of relaxation time $\gamma ^{\prime }$ in BCS which may
be suitable for low-$T_{c}$ superconductor. Instead of this, we use
the factor $k$ as fitting parameter.
\begin{figure}[tp]
\includegraphics[width=8.6cm, height=6.2cm]{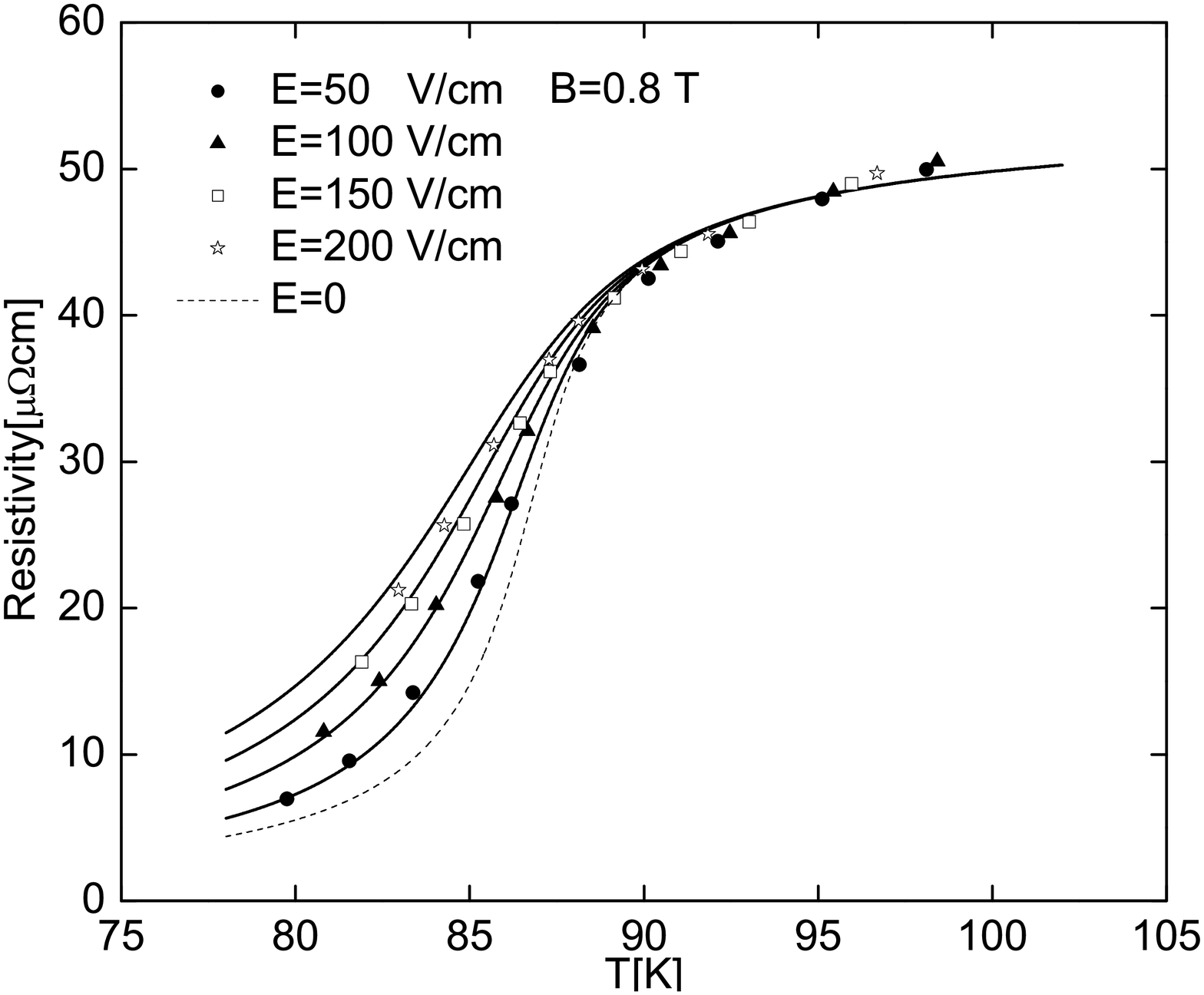}
\caption{Points are resistivity for different electric fields of an
optimally doped YBCO in Ref. \onlinecite{Puica09}. The solid line is
the theoretical value of resistivity calculated from Eq.
(\protect\ref{rho}) with fitting parameters (see text). The dashed
line is the theoretical value of resistivity in linear response with
the same parameters.} \label{fig4}
\end{figure}
The comparison is presented in Fig. 1. The resistivity
\begin{equation}
\rho =\frac{1}{\sigma _{s}+\sigma _{n}},  \label{rho}
\end{equation}%
\begin{equation}
\sigma _{s}=\frac{\sigma _{n}}{k}\overline{\sigma }_{s},
\end{equation}%
curves were fitted to Eq. (\ref{rho}) with the normal-state conductivity
measured in Ref. \onlinecite{Puica09} to be $\sigma _{n}=1.9\times 10^{4}$ $%
(\Omega $cm$)^{-1}$. The parameters we obtain from the fit are: $%
H_{c2}(0)=T_{c}dH_{c2}\left( T\right) /dT|_{T_{c}}=190$ T (corresponding to $%
\xi =13.2$ \AA ), the Ginzburg-Landau parameter $\kappa =43.6$, the order
parameter effective thickness $s^{\prime }=8.5$ \AA , and the factor $%
k=\sigma _{n}/\sigma _{GL}=0.55$, where we take $\gamma =7.8$ for optimally
doped YBCO in Ref. \onlinecite{Li02}. Using those parameters, we obtain $Gi=9.32\times 10^{-4}$ (corresponding to $\omega =0.136$%
). The order parameter effective thickness $s^{\prime }$ can be
taken to be equal to the layer distance (see in Ref.
\onlinecite{Poole}) of the
superconducting CuO plane plus the coherence length $2\xi _{c}=2\frac{\xi }{%
\gamma }$ due to the proximity effect: $3.18$ \AA $+2\frac{13.2}{7.8}$ \AA $%
=6.9$ \AA , \ roughly in agreement in magnitude with the fitting value of $%
s^{\prime }$.

We will now estimate $T_{c}^{mf}$ for this sample. For the
underdoped YBCO,
the radius of the Fermi surface of YBCO was measured in Ref. %
\onlinecite{YBCOnature}, $k_{F}=0.7$ \AA $^{-1}$, while the effective mass
is $m^{\ast }=1.9m_{e}$. We will assume that$\ $the Fermi energy for
underdoped YBCO of Ref. \onlinecite{YBCOnature} is $\varepsilon _{F}=\hbar
^{2}k_{F}^{2}/(2m^{\ast })$ and is roughly the same for the optimal YBCO
studied in this paper. The cutoff \textquotedblleft time\textquotedblright\
in physical units is then, according to Eq. (\ref{tcutoff}), $\tau
_{c}=1.39\times 10^{-17}$ s. Equation (\ref{Tmf}) gives then $%
T_{c}^{mf}=101.15$ K.
\begin{figure}[tp]
\includegraphics[width=8.6cm, height=6.2cm]{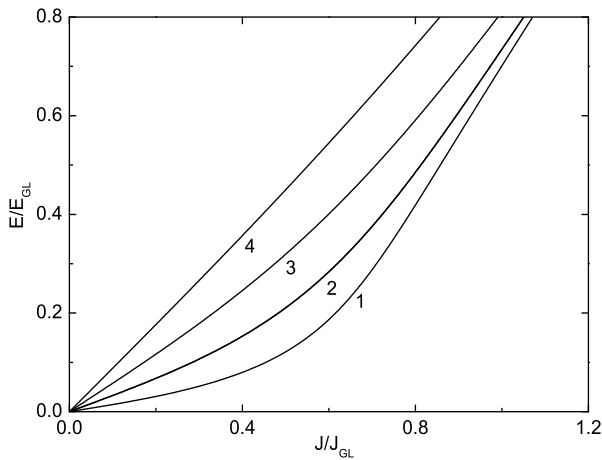}
\caption{The current-voltage curves calculated from Eq. (\protect\ref%
{finalcurrent}) by using the parameters (see text) for different magnetic
fields $b=B/Hc2$: 0.04 (1), 0.1 (2), 0.4 (3), 1.0 (4) at temperature $t=0.75$%
.}
\end{figure}
\begin{figure}[tp]
\includegraphics[width=8.6cm, height=6.2cm]{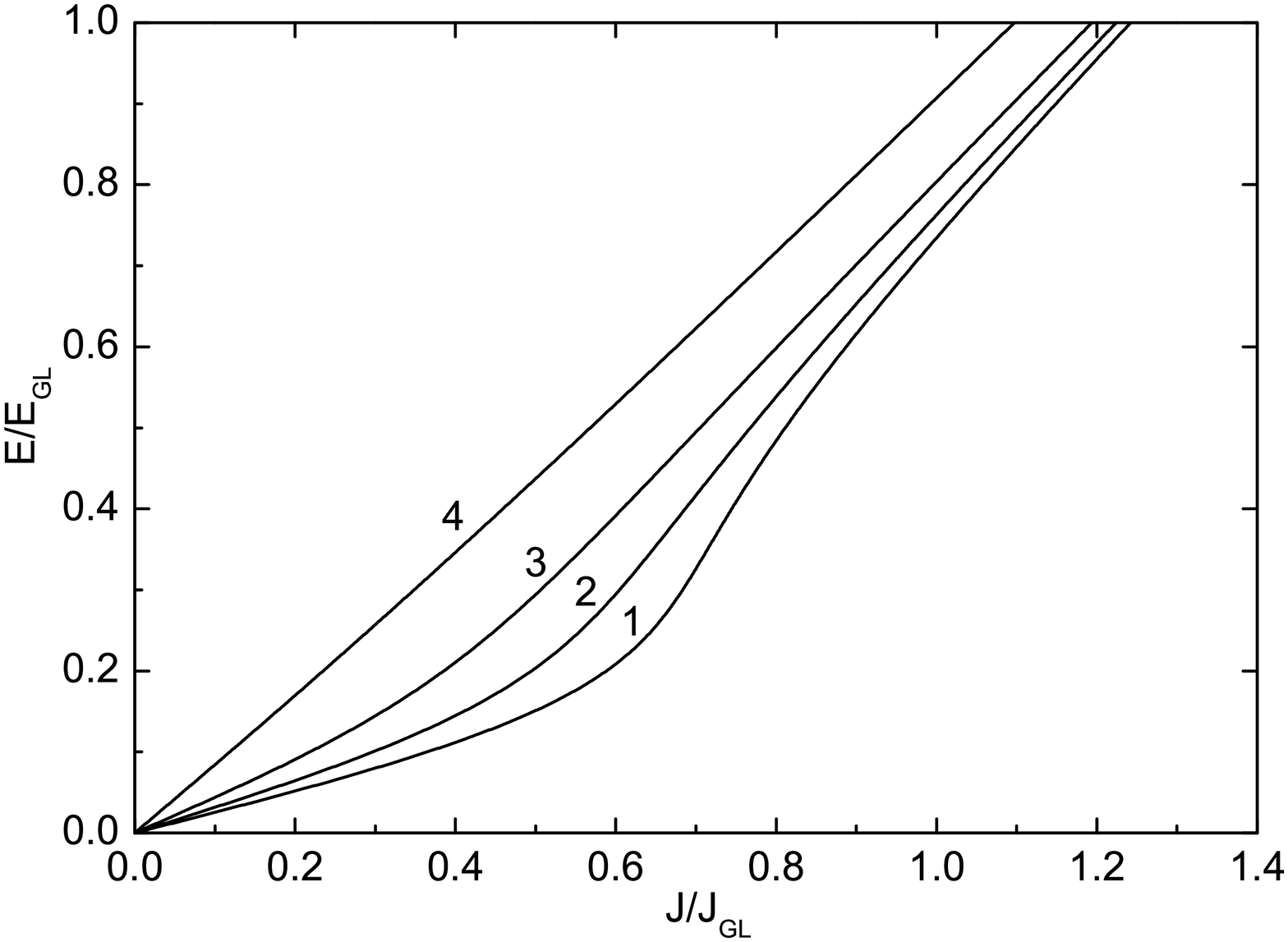}
\caption{The current-voltage curves calculated from Eq. (\protect\ref%
{finalcurrent}) by using the parameters (see text) for different
temperatures $t=T/Tc$: 0.2 (1), 0.3 (2), 0.4 (3), 1.0 (4) at magnetic flied $%
b=0.5$.}
\end{figure}
Using the parameters specified above we plot several theoretical I-V curves.
As expected the I-V curve shown in Figs. 2 and Figs. 3 has two linear
portions, the flux flow part for $E\ll E_{GL}$ and the normal Ohmic part for
$E\gg E_{GL}$. In the crossover region, $E\sim E_{GL}$, a I-V curve becomes
nonlinear due to destruction of superconductivity (the normal area inside
the vortex cores increases to fill all the space). In Fig. 2 the I-V curves
are shown for different the magnetic fields, at a fixed temperature $%
T=0.75T_{c}$. At given electric field, as the magnetic field increases, the
supercurrent decreases. When the magnetic field reaches $H_{c2}$, the I-V
curve becomes linear. In Fig. 3 the I-V curves are shown for different
temperatures, at a fixed magnetic field $H=0.5H_{c2}$. At given electric
field, as the temperature increases, the supercurrent decreases. When the
temperature reaches $T_{c}$, the I-V curve becomes linear. With decreasing
temperature the crossover becomes steeper.

\section{Discussion and conclusion}

We quantitatively studied the transport in a layered type-II superconductor
in magnetic field in the presence of strong thermal fluctuations on the
mesoscopic scale beyond the linear response. While in the normal state the
dissipation involves unpaired electrons, in the mixed phase it takes a form
of the flux flow. Time dependent Ginzburg-Landau equations with thermal
noise describing the thermal fluctuations is used to describe the
vortex-liquid regime and arbitrary flux flow velocities. We avoid assuming
the lowest Landau level approximation, so that the approach is valid for
arbitrary values the magnetic field not too close to $H_{c1}\left( T\right) $%
.

Our main objective is to study layered high-$T_{c}$ materials for which the
Ginzburg number characterizing the strength of thermal fluctuations is
exceptionally high, in the moving vortex matter the crystalline order is
lost and it becomes homogeneous on a scale above the average inter-vortex
distances. This ceases to be the case at very low temperature at which two
additional factors make the calculation invalid. One is the validity of the
GL approach (strictly speaking not far from $T_{c}\left( H\right) $) and
another is effect of quenched disorder. The later becomes insignificant at
elevated temperature due to a very effective thermal depinning. Although
sometimes motion tends to suppress fluctuations, they are still a dominant
factor in flux dynamics. The nonlinear term in dynamics is treated using the
renormalized Gaussian approximation. The renormalization of the critical
temperature is calculated and is strong in layered high $T_{c}$ materials.
The results were compared to the experimental data on HTSC. Our the
resistivity results are in good qualitative and even quantitative agreement
with experimental data on YBa$_{2}$Cu$_{3}$O$_{7-\delta }$ in strong
electric fields.

Let us compare the present approach with a widely used Londons'
approximation. Since we haven't neglected higher Landau levels, as very
often is done in similar studies \cite{Ullah90,RMP}, our results should be
applicable even for relatively small fields in which the London
approximation is valid and used. There is no contradiction since the two
approximations have a very large overlap of applicability regions for
strongly type-II superconductors. The GL approach for the constant magnetic
induction works for $H>>H_{c1}\left( T\right) $, while the Londons' approach
works for $H<<H_{c2}\left( T\right) $. Similar methods can be applied to
other electric transport phenomena like the Hall conductivity and thermal
transport phenomena like the Nernst effect. The results, at least in 2D, can
be in principle compared to numerical simulations of Langevin dynamics.
Efforts in this direction are under way.

\begin{acknowledgments}
We are grateful to I. Puica for providing details of experiments and to F.
P. J. Lin, B. Ya. Shapiro for discussions and encouragement, B. D. Tinh
thanks for the hospitality of Peking University and D. Li \ thanks for the
hospitality of National Chiao Tung University. This work is supported by NSC
of R.O.C. \#98-2112-M-009-014-MY3 and MOE ATU program, and D. Li is
supported by National Natural Science Foundation of China (grant number
90403002 and 10974001).
\end{acknowledgments}

\appendix

\section{Derivation of the Green function of the linearized TDGL equation}

In this appendix we outline the method for obtaining the Green function in
strong electric field for the linearized equation of TDGL (\ref{GFdef}). The
Green function is a Gaussian
\begin{equation}
G_{k_{z}}\left( \mathbf{r},\mathbf{r}^{\prime },\tau ^{\prime \prime
}\right) =\exp \left[ \frac{ib}{2}X\left( y+y^{\prime }\right) \right]
g_{k_{z}}\left( X,Y,\tau ^{\prime \prime }\right) ,  \label{Ansatz}
\end{equation}%
where
\begin{equation}
g_{k_{z}}\left( X,Y,\tau ^{\prime \prime }\right) =C_{k_{z}}(\tau ^{\prime
\prime })\theta \left( \tau ^{\prime \prime }\right) \exp \left( -\frac{%
X^{2}+Y^{2}}{2\beta }-vX\right) ,
\end{equation}%
with $X=x-x^{\prime }-v\tau ^{\prime \prime },Y=y-y^{\prime },\tau ^{\prime
\prime }=\tau -\tau ^{\prime }.$ $\theta \left( \tau ^{\prime \prime
}\right) $ is the Heaviside step function, $C$ and $\beta $ are coefficients.

Substituting the Ansatz (\ref{Ansatz}) into Eq. (\ref{GFdef}), one obtains
following conditions condition:
\begin{equation}
\varepsilon -\frac{b}{2}+\frac{\nu ^{2}}{2}+\frac{1}{d^{2}}[1-\cos (k_{z}d)]+%
\frac{1}{\beta }+\frac{\partial _{\tau }C}{C}=0,  \label{cond1}
\end{equation}%
\begin{equation}
\frac{\partial _{\tau }\beta }{\beta ^{2}}-\frac{1}{\beta ^{2}}+\frac{b^{2}}{%
4}=0.  \label{cond2}
\end{equation}%
The Eq. (\ref{cond2}) determines $\beta $, subject to an initial condition $%
\beta (0)=0$,
\begin{equation}
\beta =\frac{2}{b}\tanh \left( b\tau ^{\prime \prime }/2\right) ,
\label{beta.fun}
\end{equation}%
while Eq. (\ref{cond1}) determines $C$:
\begin{eqnarray}
C &=&\frac{b}{4\pi }\exp \left\{ -\left( \varepsilon -\frac{b}{2}+\frac{v^{2}%
}{2}+\frac{1}{d^{2}}[1-\cos(k_{z}d)]\right) \tau ^{\prime \prime }\right\}
\notag \\
&\times &\sinh ^{-1}\left( \frac{b\tau ^{\prime \prime }}{2}\right) .
\label{C.fun}
\end{eqnarray}%
The normalization is dictated by the delta function term in definition of
the Green's function Eq. (\ref{GFdef}).

\section{Comparison with thermodynamics}

From TDGL, we obtain in the case $\upsilon =0$:
\begin{equation}
\left\langle \left\vert \psi _{n}(\mathbf{r},\tau )\right\vert
^{2}\right\rangle =\frac{\omega tb}{2\pi s}\int_{\tau =\tau _{c}}^{\infty }%
\frac{\exp \left\{ -\left( 2\varepsilon -b+\frac{2}{d^{2}}\right) \tau
\right\} I_{0}\left( \frac{2\tau }{d^{2}}\right) }{\sinh (b\tau )}.
\end{equation}%
The superfluid density at $b=0$ and $\varepsilon =0$ can be obtained by
taking $b$ and $\varepsilon $ to zero limit in the above equation:
\begin{equation}
\left\langle \left\vert \psi _{n}(\mathbf{r},\tau )\right\vert
^{2}\right\rangle =\frac{\omega t}{2\pi s}\int_{\tau =\tau
_{c}/d^{2}}^{\infty }\frac{\exp \left\{ -2\tau \right\} I_{0}\left( 2\tau
\right) }{\tau }.
\end{equation}%
Performing the integration by parts, one obtains
\begin{equation}
\left\langle \left\vert \psi _{n}(\mathbf{r},\tau )\right\vert
^{2}\right\rangle \simeq -\frac{\omega t}{2\pi s}\{\ln \left( \tau
_{c}/d^{2}\right) +\gamma _{E}\}+O\left( \tau _{c}\right) .  \label{dynamic}
\end{equation}

In the case without external electric field (or $v=0$), the equation
obtained from TDGL shall approach the thermodynamics result. In
thermodynamics method, we shall evaluate the partition function
$Z=\int D\psi _{n}D\psi _{n}^{\ast }e^{-F_{GL}/T}$ where $F_{GL}/T$
is defined in Eq. (\ref{Boltz}). The superfluid density in the
thermodynamic approach at the phase transition point
\begin{eqnarray}
\left\langle \left\vert \psi _{n}(\mathbf{r},\tau )\right\vert
^{2}\right\rangle &=&\frac{\omega td}{\left( 2\pi \right) ^{3}s}%
\int_{0}^{k_{\max }}d\mathbf{k}  \notag \\
&\times &\int_{0}^{2\pi /d}dk_{z}\frac{1}{\frac{\mathbf{k}^{2}}{2}+\frac{%
1-\cos \left( k_{z}d\right) }{d^{2}}}  \notag \\
&\simeq &\frac{\omega t}{2\pi s}\{\ln \Lambda +\ln \left(
2d^{2}\right)\}+O\left( \Lambda ^{-1}\right),\nonumber \\
\label{thermodynamic}
\end{eqnarray}%
where $\Lambda =k_{max}^{2}/2$.

The relation between the cutoff \textquotedblleft time" $\tau _{c}$ and
energy UV cutoff $\Lambda $ is obtained by comparing Eq. (\ref{dynamic})
with Eq. (\ref{thermodynamic})
\begin{equation}
\tau _{c}\simeq \frac{1}{2\Lambda e^{\gamma _{E}}}.  \label{cutoffrelation}
\end{equation}

We also remark that in thermodynamic approach, if we use the
Gaussian approximation, we will get the exact same equation derived in Eq. (\ref%
{gapequation}) without electric field derived from TDGL after using Eq. (\ref%
{cutoffrelation}).

\end{document}